\documentclass[submitted,copyright]{eptcs}
\providecommand{\event}{AMMSE 2011} 
\usepackage{breakurl}             

\usepackage[scriptsize]{subfigure}
\usepackage{graphicx}
\usepackage[dvips,matrix,all]{xy}
\usepackage{array}
\usepackage{version} 

\title{Typed lambda-terms in categorical attributed graph transformation}

\author{Bertrand Boisvert, Louis F\'eraud, Sergei Soloviev
\institute{IRIT}
\institute{Universit\'e Paul Sabatier\\
Toulouse, France}
\email{\{boisvert,feraud,soloviev\}@irit.fr}
}

\def\titlerunning{Typed lambda-terms in categorical attributed graph transformation}
\def\authorrunning{B Boisvert, L. Feraud \& S. Soloviev}

\begin{document}
\maketitle

\begin{abstract}
  This paper deals with model transformation based on attributed graph rewriting. Our contribution investigates a single pushout approach for applying the rewrite rules. The computation of graph attributes is obtained through the use of typed $\lambda$-calculus with inductive types. In this paper we present solutions to cope with single pushout construction for the graph structure and the computations functions. As this rewrite system uses inductive types, the expressiveness of attribute computations is facilitated and appears more efficient than the one based on $\Sigma$-algebras. Some examples showing the interest of our computation approach are described in this paper.
\end{abstract}

\section{Introduction.}

There is currently a need of rigorous support to model based software engineering. In model-driven software engineering, models are mostly described using a graphical syntax (UML, SDL, ...). Models are composed of a structural part which can be represented as a graph and of attributes which are informations attached to vertices or edges of the graph. Thus, models can be formalized as attributed graphs and model transformation as attributed graph transformation. An attributed graph transformation is composed of a rewrite of the structural part and of computations on its attributes. Thus, we need a formal framework that can express these two types of transformation. 

Graph rewriting systems based on category theory have been widely used to deal with the transformation of structural part. 
One of the challenges of attributed graph rewriting systems concerns the
implementation of attribute computations. Most of the existing systems based on category theory 
adopt the standard algebraic approach where graphs are attributed using
algebraic data types represented by $\Sigma$-algebras~\cite{HLR, Orejas:2011:SGA:1937191.1937400}.
However, the
implementation of computations with algebraic data types meets many
difficulties, and for the sake of efficiency considerations and convenient
uses, these systems do not generally implement the whole attribute
computations but rely on programs written in a host-language~\cite{AGG_homepage}.

In our earlier work~\cite{2009.157, Maxime2008, Tran2010} we suggested
to use inductive types and lambda terms in combination with a
modification of the double pushout approach~\cite{DBLP:conf/gg/1997handbook} called DPoPb (``double
pushout-pullback'' approach). Our goal was to use the well
developed double pushout approach to implement rewriting of the
structural part of graphs and to use the expressive power of $\lambda$-terms and inductive types to describe and facilitate attribute computations. But the construction of the double pushout imposed restricting constraints on computation functions mostly due to the usage of total maps and the obligation to split computations into two parts. That is why we now present a new approach based on single pushout.

The first section of this paper introduces the main approaches of graph rewriting based on category theory, and particularly the single pushout approach on which our approach is based. Second we define our category of attributed graphs, and then explain how to apply a rewrite rule by the computation of a weak pushout. Finaly we present examples. 

\section{Categorical graph rewriting.}

\begin{figure}[h]
  \centering
  \subfigure[DPo approach]{
  $$
  \xymatrix@+1cm{
    L \ar[d]_i &K \ar[l]_l \ar[r]^r \ar[d]|-{d} \ar @{} [dl] |{(PO1)} \ar @{} [dr] |{(PO2)}&R \ar[d]^{i^{*}}\\
    G&D \ar[l]^{l^{*}} \ar[r]_{r^{*}} &H\\
  }
  $$
  \label{SchemaDPo}
  }
  \subfigure[SPo approach]{
    \centering
    $$
    \xymatrix@+1cm{
      L \ar[d]_i \ar[r]^r &R \ar[d]^{i^{*}} \ar @{} [dl] |{(PO)} \\
      G \ar[r]_{r^{*}} &H \\
    }
    $$
    \label{SchemaSPo}
  }
  \label{fig:CategoricalApproaches}
  \caption{Classical categorical graph rewriting approaches}
\end{figure}
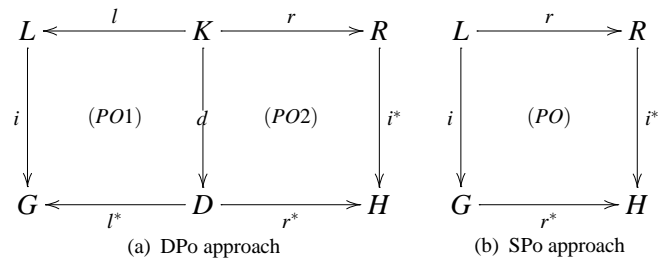

In graph rewriting systems based on category theory, we usualy define a category whose objects are graphs and morphisms are graph homomorphisms. A transformation rule is composed of at least two graphs called the left-hand side (usually noted $L$) and right-hand side (usually noted $R$). The left-hand side
describes which subgraph a graph $G$ must contain in order that
the transformation could be applied to it, and the right-hand side describes how this part
will look like after the transformation. Morphisms between left-hand side and right-hand side describe which parts of graphs will be deleted, transformed or added.
To apply a rule to some
subgraph of a larger graph $G$, we need first
to embed the left-hand side as a subgraph of $G$.
The embedding is represented by an inclusion $L\stackrel{i}{\rightarrow} G$. Cf Figure \ref{SchemaDPo} and \ref{SchemaSPo}.

There are two principal categorical approaches to graph rewriting: double pushout (abbreviated DPo, concieved by H. Ehrig and his colleagues~\cite{DBLP:conf/gg/Ehrig78}, \cite{DBLP:conf/gg/1997handbook}) and single pushout (abbreviated SPo, mainly developped by L\"owe~\cite{SPoLowe}, \cite{DBLP:conf/gg/1997handbook}). The main difference is that in DPo morphisms are total maps and in SPo morphims are partial maps. This implies different forms of rules.

In the DPo approach a rule is defined by 3 graphs and 2 total morphisms: $L\stackrel{l}{\leftarrow} K\stackrel{r}{\rightarrow} R$. The morphism l indicates what vertices or edges should be erased (the ones who are not in the image of $l$) and the morphism $r$ indicates what vertices or edges should be transformed (those who are in $K$), and added (those who are not in the image of $r$). The application of the rule is done by a computation of a pushout-complement (adding the arrows $K\stackrel{d}{\rightarrow}D$ and $D\stackrel{l^*}{\rightarrow} G$ and then a pushout (the arrows $R\stackrel{i^*}{\rightarrow} H$ and $D \stackrel{r^*}{\rightarrow} H$). Cf Figure \ref{SchemaDPo}.

In the SPo approach, a rule is defined by one partial morphim $L\stackrel{r}{\rightarrow} R$.
Vertices and edges not included in the domain of $r$ will be deleted, the ones in the domain of $r$ will be transformed and those which are not in the image of $r$ will be added. The application of the rule is done by the computation of one pushout (adding the arrows $G\stackrel{r^*}{\rightarrow} H$ and $R\stackrel{i^*}{\rightarrow} H$). Cf Figure \ref{SchemaSPo}

Because not all pushout-complements necessarily exist in the categories of graphs, there exist ``application conditions'' in DPo approach. As a consequence, rules that create dangling edges are forbidden in the DPo approach while in SPo approach dangling edges are removed when the rule is applied. If necessary, it is possible to add application conditions in the SPo approach as well.
Thus the SPo approach is more general than the DPo approach, but SPo approach remained less developed due, in our opinion,  
mostly to historical reasons and to the fact that computation of
pushout in categories of partial maps is more difficult than in
categories of total maps.

Both approaches met many difficulties on the level of
attribute computations. Our experience with the DPoPb approach \cite{Maxime2008, 2009.157, Tran2010} and the usage of $\lambda$-terms for attributes was encouraging but the construction of a double pushout imposed us some constraints due to the use of total maps and the obligation to split computation into two parts. The new approach we present here is more direct and natural, free of application conditions and we have no more constraints on the computational level. The rewriting process involves structure transformation and attribute computations. In this paper, we would stress on attribute computations and lighten the structure rewrite. We hope that the combined use of SPo and $\lambda$-terms will permit to overcome many known difficulties of attribute computations.

To develop a categorical graph rewriting system we must define a category (objects and morphisms) and then explain how to apply a rule (in our case by the computation of a pushout).

A pushout of two
morphisms $L\stackrel{r}{\rightarrow} R$, $L \stackrel{i}{\rightarrow} G$ is a couple of morphisms ($G \stackrel{r'}{\rightarrow}H$, $R \stackrel{i'}{\rightarrow}H$) such that:
\begin{itemize}
  \item $i' \circ r = r' \circ i$
  \item for every other couple of morphisms ($R \stackrel{h}{\rightarrow}H'$, $G \stackrel{g}{\rightarrow}H'$) such that $h \circ r = g \circ i$ it exists a unique morphism $c$ such that the diagram below commutes:
\end{itemize}

$$
\xymatrix@+1cm{
L \ar[d]_i \ar[r]^r &R\ar[d]^{i'}\ar[ddr]^h&\\
G\ar[drr]_g\ar[r]^{r'}&H\ar@{-->}[dr]^c&\\
&&H'\\
}
$$

As a consequence, the existence of pushout
implies the uniqueness of the object $H$ up to isomorphism
(cf.~\cite{fundamentals_alg_graph_transf, SPoLowe}). If we have the two properties in the definition of pushout but not the unicity of $c$, the construction is called a weak pushout.

\section{Category of Attributed Graphs.}
\label{CategoryAttributedGraphs}

We shall denote $Gr^T$ the category of graphs we
consider below.\\

{\bf Objects.} Objects of $Gr^T$ are oriented attributed
graphs. We shall assume that the vertices and edges are sets (noted $V(G)$ and $E(G)$ for a graph $G$) of natural numbers with $V(G)\cap E(G)= \emptyset$ and that a standard (lexicographic) ordering on vertices and edges is defined and noted $<$.
This will help us to avoid ambiguity in the definition of
morphisms, and in any case this assumption is standard when
implementations are considered. The attributes will be $\lambda$-terms.  The system of lambda calculus in this paper is the simply
typed lambda calculus with surjective pairing, terminal object and
inductive types, (see for example~\cite{DCH}). We shall
denote the function type by $A\rightarrow B$ and 
the product type for types $A$ and $B$ by $A\times B$. The presence of inductive
types permits to define all ordinary types of attributes, like
$Bool$, $Nat$, etc., as well as more complex types like lists,
binary trees, $\omega$-trees, etc. We prefer to include pairings in the syntax directly instead of defining it using inductive types.

We decide to have exactly one attribute per node or edge. Pairing permits to embed different datas into this unique attribute. We can see each attribute as a tuple containing all informations attached to a node or an edge. The n-tuple
$<t_1,..., t_n>$ is considered as an abbreviation
of the term $<...<t_1, t_2>,...,t_n>$. If $A_1,..., A_n$ are
types of $t_1,..., t_n$ respectively, the type of this tuple will be written
as $A_1\times...\times A_n$ instead of $(...(A_1\times A_2)\times... A_n)$.
We shall use trivial attribute $0:T$ ($T$ - terminal object)
to represent the absence of attributes.
Thus we have a bijection between the set of nodes and edges and the set of attributes which permits to simplify some proofs.

If $G$ is an attributed graph, $V(G)$ is the set of its
vertices, $E(G)$ the set of its edges, $att(v)$ where $v\in V(G)\cup E(G)$
is the corresponding attribute ($\lambda$-term).\\

\begin{figure}
  \centering
  \subfigure[Structural part]{
    \includegraphics[width=0.5\textheight]{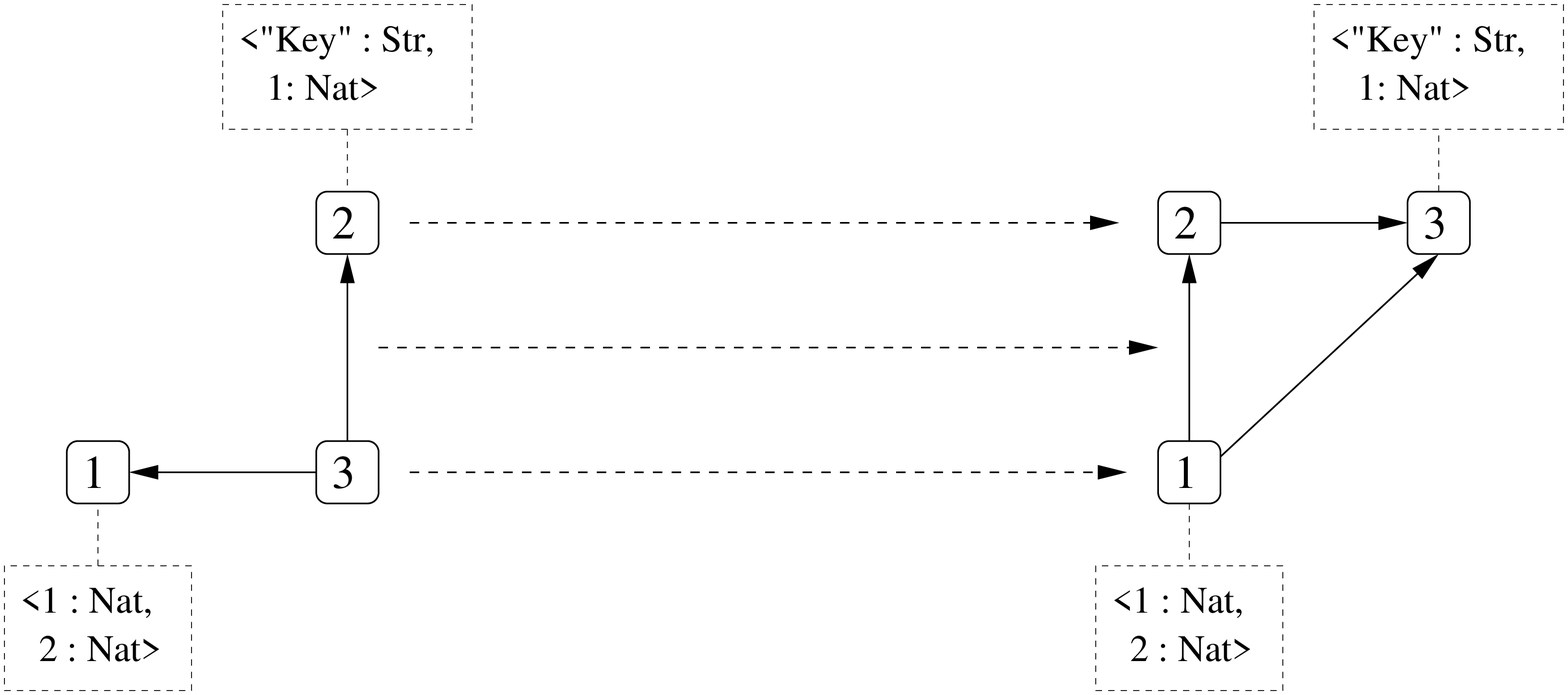}
    \label{fig:morphismeGraphesStructure}
  }

  \subfigure[Structural part + attribute dependency relation]{
    \includegraphics[width=0.5\textheight]{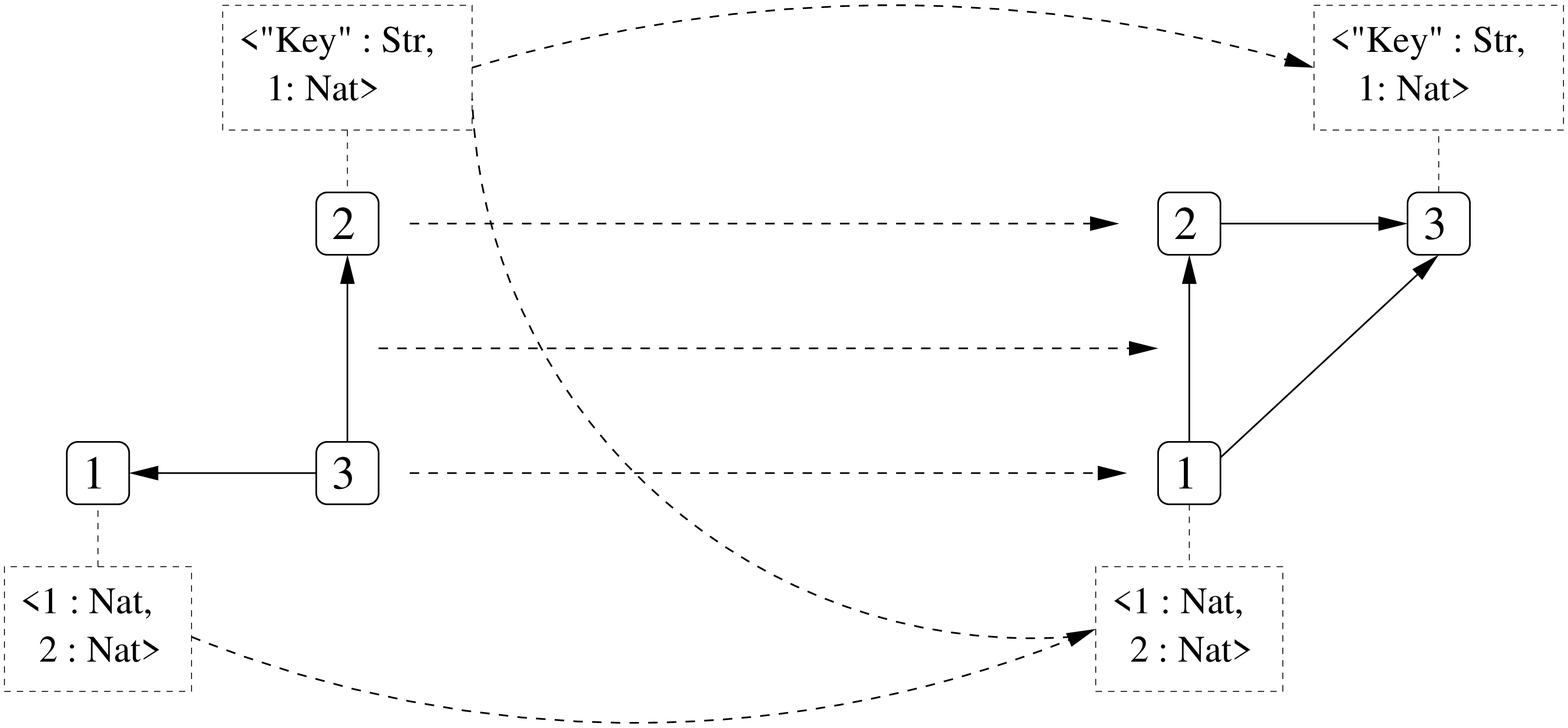}
    \label{fig:morphismeGraphesStructureTypes}
  }

  \subfigure[Structural part + attribute dependency relation + computational part]{
    \includegraphics[width=0.5\textheight]{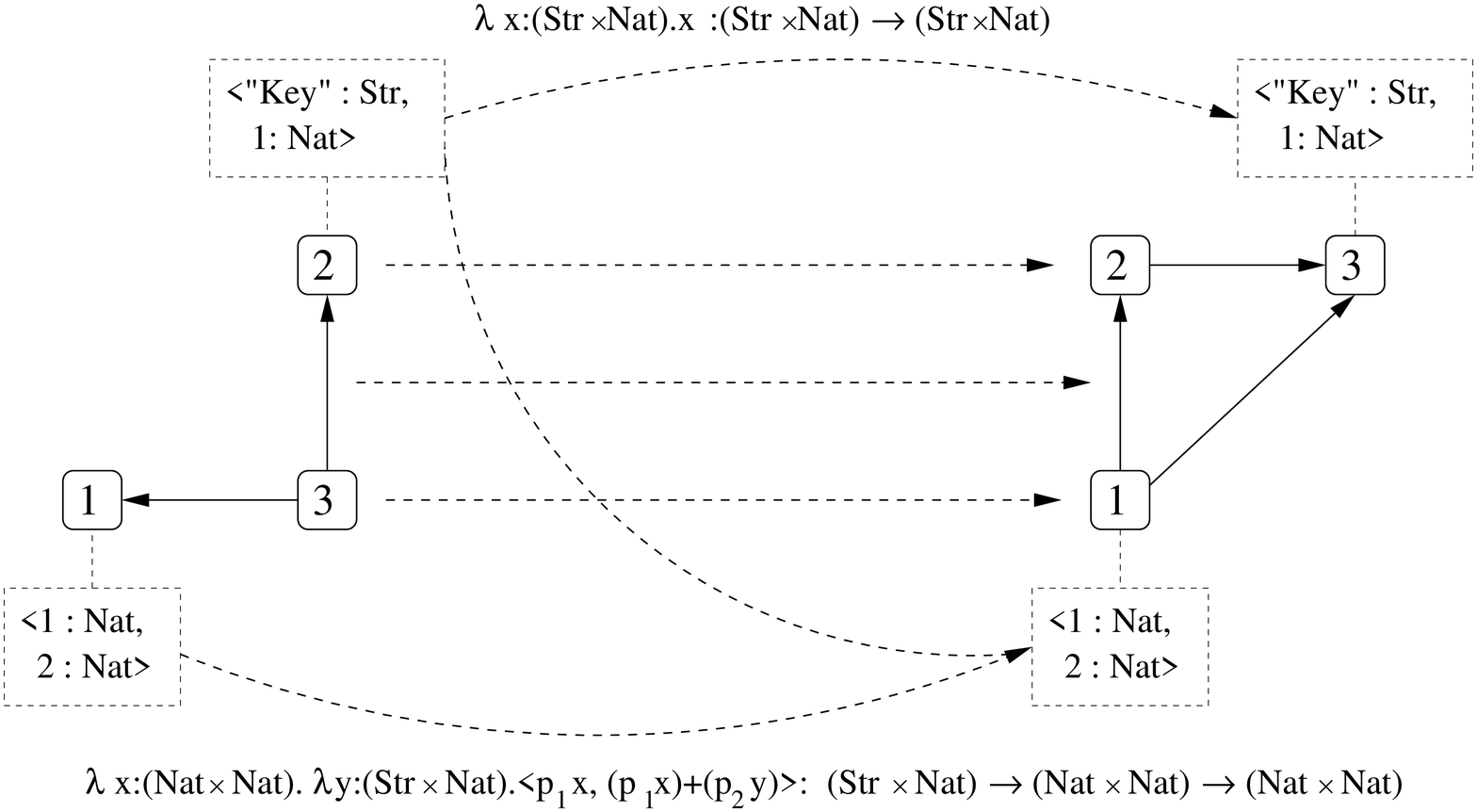}
    \label{fig:morphismeGraphesStructureTypesTermes}
  }

  \caption{Three-level attributed graph morphism}
  \label{fig:morphismeGraphes}
\end{figure}

{\bf Three-level morphisms.} Let $G, H$ be two attributed graphs.
We shall assume that all $\lambda$-terms considered below are
typed in the same context $\Gamma$. This context may be fixed
for the whole category, or at least sufficient for all graphs
and terms in consideration. The terms are not necessarily closed.
The equality of terms is understood
in ordinary sense as equality w.r.t. $\alpha, \beta, \eta$ and also
$\iota$ conversion for recursion\footnote{In principle, 
other forms of equality could be considered.}.
Morphisms $f:G\rightarrow H$ are defined using
the following three-level construction:

\begin{enumerate}
\item The ``structural part'' $f_{str}$ is a partial graph homomorphism
(without attributes) of $G$ to $H$ (cf. Fig. \ref{fig:morphismeGraphesStructure}).
\item The ``attribute dependency relation'' $f_{adr}$ is a relation
between the sets
$ V(G)\cup E(G)$ and $V(H)\cup E(H)$ induced by computation functions. For each $v\in  V(H)\cup E(H)$
its preimage (i.e. the set of all its antecedents) is $[v]_{f_{adr}}\subseteq  V(G)\cup E(G)$ and represents all attributes of graph $G$ that we can use to compute $v$.
(We could consider here instead of vertices and edges corresponding
attributes, cf. Fig. \ref{fig:morphismeGraphesStructureTypes}.)
\item The ``computational part'' is represented, by the $\lambda$-term $f_{cmp}(v)$,
for each element $v\in  V(H)\cup E(H)$.
These $\lambda$-terms will be called computation
functions. They are functions that take as argument attributes of graph $G$ defined by $f_{adr}$.
More precisely, let $v\in  V(H)\cup E(H)$. Let
$att(v) = t:A$ be the corresponding attribute.
Let $[v]_{f_{adr}}=\{u_1,...,u_k\}, u_1<...<u_k$ (we use here the fact that the
vertices are natural numbers) and
$$att(u_1) = t_{1}:A_{1},..., att(u_k) = t_{k}:A_{k}$$
be the attributes of the antecedents.
Now the term $s_v=f_{cmp}(v)$ has the type
$$A_{1}\rightarrow ... \rightarrow A_{k}\rightarrow A$$
and should
satisfy the following property:
$$s_v(t_{1},...,t_{k}) = t$$
(of course, many arguments may be ``dummy'').
In particular, if $[y]_{f_{adr}}=\emptyset$ then $s_v=t$.
(See Fig. \ref{fig:morphismeGraphesStructureTypesTermes}.)
\end{enumerate}

{\bf Equality of morphisms.}
Two morphisms $f, g:G\rightarrow H$ are equal if
\begin{enumerate}
\item $f_{str}=g_{str}$;
\item the relations $f_{adr} = g_{adr}$;
\item and for each $v\in V(H)\cup E(H)$ their computation functions $f_{cmp_v}$ and $g_{cmp_v}$ are equals on their arguments in $G$. More precisely, let $[v]_{f_{adr}} = [v]_{g_{adr}} = \{u_1,...,u_k\}$ be the arguments of $f_{cmp_v}$ and $g_{cmp_v}$; we have:
  $$f_{cmp_v}(u_1,...,u_n)=g_{cmp_v}(u_1,...,u_n)$$ as $\lambda$-terms.
\end{enumerate}

{\bf Remarks: }
\begin{itemize}
\item One may notice that the equality of morphisms does not imply the equality of computations functions and two morphisms can be equal and have different computation functions.
\item If the attributes of $G$ are variables then the equality of functions on values is equivalent to the equality of computation functions. (This choice of variables as attributes is natural if we consider rule schemes instead of instance of the rules).
\item Given two morphisms $f,g: G \rightarrow H$, the equality of first two levels $f_{str}= g_{str}$ and $f_{adr}=g_{adr}$ implies $f=g$ because the values of the attributes of H are the same.
\item Taking into account the way how the rules are applied, it is in good accord with the intuition. A mosphism $r: L \rightarrow R$ is used for the formulation of a rule (or rule scheme) and it is natural that the values (attributes of $R$) are known. The computation occurs in the application of a rule (construction of a pushout).
\end{itemize}

{\bf Composition.}
\begin{enumerate}
\item On the level of structure, we take the composition $g_{str}\circ f_{str}$;
\item On the level of attribute dependency relations we take the
composition of relations $(g_{adr}\circ f_{adr})$ .
One may note that:
$$[w]_{(g\circ f)_{adr}}=[w]_{g_{adr}\circ f_{adr}} = \cup_{v\in [w]_{g_{adr}}} [v]_{f_{adr}}$$
\item On the level of computation functions ($\lambda$-terms) the
composition is defined using composition of $\lambda$-terms.

More
precisely, let:
\begin{itemize}
\item $t = g_{cmp}(w)$;
\item $[w]_{g_{adr}} =\{v_1,..., v_k\},
v_1<...< v_k$;
\item $t_1=f_{cmp}(v_1),..., t_k=f_{cmp}(v_k)$;
\item $[v_1]_{f_{adr}} = \{u_{11},..., u_{1n_1}\},..., [v_k]_{f_{adr}} =
\{u_{k1},..., u_{kn_k}\}$.
\end{itemize}

The intersections of the
antecedents may be not empty, so let $u_1< ... < u_p$ be the
distinct elements (vertices or edges) of the union $[v_1]_{f_{adr}}\cup...\cup [v_k]_{f_{adr}}$.
Let $A_1,..., A_p$ be the types of attributes $att(u_1),..., att(u_p)$
respectively, and $x_1:A_1,..., x_p:A_p$ term variables not belonging to the context $\Gamma$. Since each of the elements
$u_{ij}$ corresponds to exactly one of $u_1,..., u_p$, we
have $u_{ij}=u_m$ for some $m, 1\leq m\leq p$, and for
term variables we may put $x_{ij} = x_m$.

Now we define:
$$(g\circ f)_{cmp}(w) =_{df} \lambda x_1:A_1...\lambda x_p:A_p.(t(t_1x_{11}...
x_{1n_1})...(t_kx_{k1}...x_{kn_k})).$$

\end{enumerate}

{\bf Identity morphisms.}
For an attributed graph G, we call $Id_G$ its identiy morphism defined by:
\begin{enumerate}
\item for the structural part, we take the identity graph homomorphism;
\item for the attribute dependency relation, we take the identity relation;
\item for the computation functions, for each $v\in V(G)\cup E(G)$, let $A$ be the type of $att(v)$, $(Id_G)_{cmp}(v)=\lambda x:A.x : A \rightarrow A$.\\
\end{enumerate}

{\bf Theorem.} Attributed graphs and graph morphisms described above form
a category.\\

{\bf Proof:}
Composition is associative due to associativity of the composition of graph homomorphisms, and associativity of the composition of relations. For $\lambda$-terms composition is associative too because of confluence and the fact that all simply typed $\lambda$-terms are strongly normalizable. Thus any evaluation strategy will terminate on a same simply typed $\lambda$-term.
It is easy to verify that for every morphism $f: G \rightarrow H$ we have $f \circ Id_G = f$ and $Id_H \circ f = f$.

\section{Weak pushout computation in category $Gr^T$.}

As we said, to apply a rule $L \stackrel{r}{\rightarrow}R$ to a graph $G$ we must find an embedding $R \stackrel{i}{\rightarrow}G$ and then compute the pushout of $r$ and $i$. But in our attributed graphs, the attributes of $R$ may contain free variables that are instanciated in $G$. So we shall see r as a rule ``scheme''. 
The application consists in general in two steps. First we take an instance of $L \stackrel{r}{\rightarrow}R$ obtained by substitution of $\lambda$-terms for certain free variables. Afterwards we try to ``embed'' the left side into $G$.\\

{\bf Injective attributed graph morphism:}
Let $f:G \rightarrow H$ be an attributed graph morphism.
$f$ is injective if:
\begin{enumerate}
\item $f_{str}$ is an injective partial graph homomorphism (i.e. $\forall v_1,v_2 \in V(G)\cup E(G).(f_{str}(v_1) = f_{str}(v_2) \Rightarrow v_1 = v_2)$);
\item $f_{adr} = f_{str}$;
\item for all $v'\in V(H)\cup E(H)$:
  \begin{itemize}
    \item if $[v']_{f_{adr}}$ is empty, $f_{cmp}(v')=att(v')$,
    \item if $[v']_{f_{adr}}$ is not empty (thus $[v']_{f_{adr}}$ is a singleton that we denote $\{v\}$ because $f_{adr}$ is injective), $f_{cmp}(v')=\lambda x:A.x : A \rightarrow A$ where A is the type of $att(v)$.
  \end{itemize}
\end{enumerate}

\bigskip

{\bf Canonical retraction of a total injective attributed graph morphism:}
Let $f:G \rightarrow H$ be a total injective attributed graph morphism.
A retraction of $f$ (or a left inverse) is an attributed graph morphism $\overline{f}: H \rightarrow G$ such that $\overline{f} \circ f = Id_G$.

With this definition, we have not necessarily $f \circ \overline{f} = Id_H$, and $\overline{f}$ is not unique in general. That's why we give a canonical construction to obtain a retraction of $f$. This construction is defined by:

\begin{enumerate}
\item for every $v'\in V(H)\cup E(H)$ if $[v']_{f_{str}}$ is empty, $v'$ has no image by $\overline{f}_{str}$; if $[v']_{f_{str}}$ is not empty (thus $[v']_{f_{str}}$ is a singleton that we denote $\{v\}$ because $f_{str}$ is injective), $\overline{f}_{str}(v') = v$.
\item $\overline{f}_{adr} = \overline{f}_{str}$
\item for each $v\in V(G)\cup E(G)$:
  \begin{itemize}
    \item if $[v]_{\overline{f}_{adr}}$ is empty, $\overline{f}_{cmp}(v)=att(v)$;
    \item if $[v]_{\overline{f}_{adr}}$ is not empty (thus $[v]_{\overline{f}_{adr})}$ is a singleton that we denote $\{v'\}$), $\overline{f}_{cmp}(v)=\lambda x:A.x : A \rightarrow A$ where A is the type of $att(v')$.
  \end{itemize}
\end{enumerate}
As $f$ is a total injection, it is easy to see that $\overline{f} \circ f = Id_G$.\\

{\bf Construction of a weak pushout}
The construction of a (weak) pushout in case of application
of a rule is inspired by the paper by L\"owe and others~\cite{DBLP:conf/gg/1997handbook}, but there will be differences due to
our definition of attributed graphs and graph morphisms.

The ``starting point'' is the pair of morphisms
($L\stackrel{r}{\rightarrow} R$, $L\stackrel{i}{\rightarrow} G$) 
where $i$ is injective and total attributed graph morphism as definded above. We want to compute the weak pushout ($R\stackrel{i'}{\rightarrow} H$, $G\stackrel{r'}{\rightarrow} H$) of this pair.

The definition of pushout in the paper by  L\"owe \cite{DBLP:conf/gg/1997handbook} uses
coequalizers. We will have in mind this construction,
but will give here a more restricted and straightforward definition,
without detailed study of coequalizers in the category $Gr^T$.

First step to define a pushout using coequalizers in the category
of graphs would be to take the coproduct $G+R$ of $G$ and $R$
(coproduct being here just the disjoint union). Next step
would be to factorize it by certain equivalence relation (creating $(G+R)'$ which contains equivalence classes),
and then to complete the construction using composition
with certain morphism $p$ from factor object to pushout object
$H$.

\begin{figure}
$$
\xymatrix@+0.5cm{
L \ar@<-2pt>[d]_i \ar[r]^r &R\ar@<2pt>[dr]^{j''} &\\
G \ar@<-2pt>[u]_{\overline{i}} \ar@<-2pt>[dr]_{j'}&&G+R\ar[dr]^{f''} \ar@<2pt>[ul]^{\overline{j''}}&\\
&G+R\ar[dr]_{f'} \ar@<-2pt>[ul]_{\overline{j'}} &&(G+R)'\ar[d]^p\\
&&(G+R)'\ar[r]_p&H\\
}
$$
\caption{Construction of weak pushout}
\label{fig:constructionWeakPushout}
\end{figure}
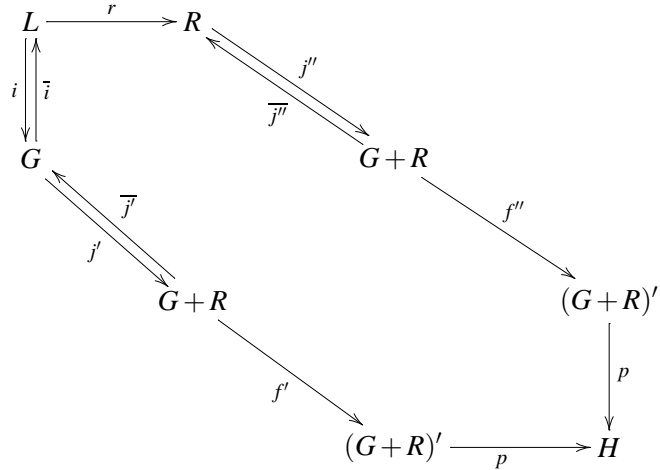

We shall define each of the morphisms $r'$ and $i'$ as a composition
of three morphisms (Cf figure \ref{fig:constructionWeakPushout}) in order
to have
$$r' = G\stackrel{j'}{\rightarrow}(G+R)\stackrel{f'}{\rightarrow}(G+R)'\stackrel{p}
{\rightarrow}H$$ and
$$i' = R\stackrel{j''}{\rightarrow}(G+R)\stackrel{f''}{\rightarrow}(G+R)'\stackrel{p}
{\rightarrow}H$$

The objects and morphisms in these
diagrams are defined in several steps.

\begin{itemize}
\item On the level of structure $G+R$ is disjoint union
of the graphs $G$ and $R$;
\item on the level of attributes each element of $G$
and $R$ in $G+R$ has the same attribute as in
$G$ and $R$;
\item $j'$ and $j''$ are inclusions
 respectively of $G$ and $R$ into $G+R$, thus they are total injective attributed graph morphisms.
\end{itemize}

To continue, we define first the equivalence relation
$\sim_1$ on the elements of the graph structure $G+R$.
\begin{itemize}
\item let's put $a\sim_1 b$ for $a, b\in G+R$ if
$\exists x\in L. (j'(i(x))=a \wedge j''(r(x))=b)$
\item then the relation $\sim$ is defined as
reflexive, symmetric and transitive closure
of $\sim_1$.
\item notice that the elements of $G+R$ which
are not the images of elements of $G-i(dom(r))$ form
equivalence classes consisting of single
element (itself).
\end{itemize}

The elements of $(G+R)'$ are defined as equivalence classes
of elements of $G+R$. It is easily checked that
this definition is consistent with the incidence
relation and the map sending each element of $G+R$
to its equivalence class is a (total) graph homomorphism.
This map will be structural part of $f'$ and $f''$.

Moreover, each equivalence class with respect to $\sim$
containing an image of an element of $R$ 
may be seen as a ``span'', consisting of the image of this element
of $R$ under $j''$ and the images of its antecedent via $r$
under $j'\circ i$. In particular, each equivalence class
contains exactly one image of an element of $R$.
As a consequence, the composition $f''_{str}\circ j''_{str}$
is injective.

It permits also to define the attribute part of $(G+R)'$.
Each equivalence class that contains an image of an
element of $R$ has the same attribute as this element
has in $R$. Other equivalence classes (that have
the form $\{j'(y)\}, y\in G, y\neq i(x)$ for some
$x\in L$) keep the same attribute as in $G$.

The definitions of relational part and computation functions
of $f'$ and $f''$ are different.

For $f''$ the relation $f''_{adr}$ connects the elements of $R$ with
corresponding equivalence classes (it is bijective on the
$R$-part). There is no connections on the $G$-part. The
computation functions are identities.

{\bf Remark.} The composition $f''\circ j''$ is injective, in
particular $(f''\circ j'')_{str}$ is an injective total graph
homomorphism, $(f''\circ j'')_{adr}=(f''\circ j'')_{str}$ and
computation functions are identities.

Now we may define $f'$ as follows:
\begin{itemize}
\item $\forall v \in img(j' \circ i)$:
 $$f'=
G+R\stackrel{\overline{j'}}{\rightarrow}G\stackrel{\overline{i}}{\rightarrow}
L\stackrel{r}{\rightarrow}R\stackrel{j''}{\rightarrow}G+R\stackrel{f''}{\rightarrow}
(G+R)'.$$
\item for the elements of $G - i(L)$, $f'$ is like the identity.
\end{itemize}

As usual (cf. \cite{DBLP:conf/gg/1997handbook}) $H$ is defined now as for coequalizer
construction. Let $L_0=dom(r)$. In our case $H$ will be a subgraph
of $(G+R)'$. The incidence relation in $(G+R)'$ is inherited from $R$ and $G$. The elements on H (on the level of graph structure) are:
\begin{enumerate}
\item all the equivalence classes of the form $\{x_1,..., x_k,
z\}$ ($x_1,..., x_k\in j'(i(L_0)),  z\in j''(r(L))$;
\item all the equivalence classes of the form $\{z\}$, $z\in j''(R-r(L))$;
\item all the equivalence classes of the form $\{x\}$, $x\in
j'(G-i(L))$ that are not dangling edges \cite{DBLP:conf/gg/1997handbook}.
\end{enumerate}

The
attributes for the equivalence classes of the first two types are
inherited from $R$ and for the third from $G$.

The morphism $p$ is defined as follows. Its structural part is
identity on all elements of $(G+R)'$ that remain in $H$. 
We have also $p_{adr}=p_{str}$,
and all computation functions are identities.

Now $i'$, $r'$ and $H$ are defined such that $i' \circ r = r' \circ i$.

\begin{figure}
\centering
\begin{tabular}{cc}
$
\xymatrix@+1cm{
L \ar[d]_i \ar[r]^r &R\ar[d]^{i'}\ar[ddr]^h&\\
G\ar[drr]_g\ar[r]^{r'}&H\ar@{-->}[dr]^c&\\
&&H'\\
}
$ & \begin{minipage}{3cm}$r' = p \circ f' \circ j'$\\ $i'= p \circ f'' \circ j''$ \end{minipage}\\
\end{tabular}
\caption{Definition of Pushout}
\label{schemaPo}
\end{figure}
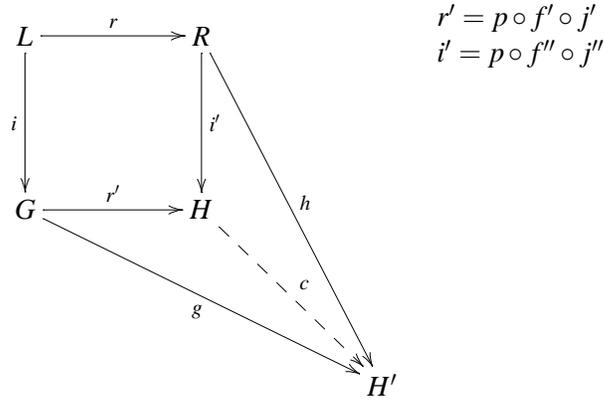

Let $h:R \rightarrow H'$ and $g: G \rightarrow H'$ be two other morphisms such that $h \circ r = g \circ i$. As $i'$ is injective, we can use the canonical retraction $\overline{i'}$ and take for $c$ (cf figure \ref{schemaPo}) $h \circ \overline{i'}$ and for elements who are not in the domain of $\overline{i'}$ we extend $c$ in order to make it in accord with $g$. The commutativity on the level of computation functions follows from the definition of equality of attributed graph morphisms (cf section \ref{CategoryAttributedGraphs}). Thus the diagram commutes but in general the unicity of $c$ is not guaranteed, so we have a weak pushout.

\section{Examples.}

To illustrate our transformation approach we present in this section two examples: the first one compares the graph grammar to compute $n!$ in our framework with the graph grammar given on the $AGG$ website \cite{AGG_homepage}, i.e. the graph grammar corresponding to the framework based on $\Sigma$-algebras. And the second one presents computation on attributes representing infinite trees which is not possible by using $\Sigma$-algebras. Let us mention also some examples that we do not develop in this paper but that can be easily developped using our approach:
\begin{itemize}
\item graph cloning \cite{SPb2011};
\item information balance between attributes and structure;
\item computation on functional attributes.
\end{itemize}

\subsection{Computation of $n!$.}
\begin{figure}[h!]
  \centering
  \subfigure[$n!$ computation based on $\Sigma$-algebras]{
    \includegraphics[width=0.41\textwidth]{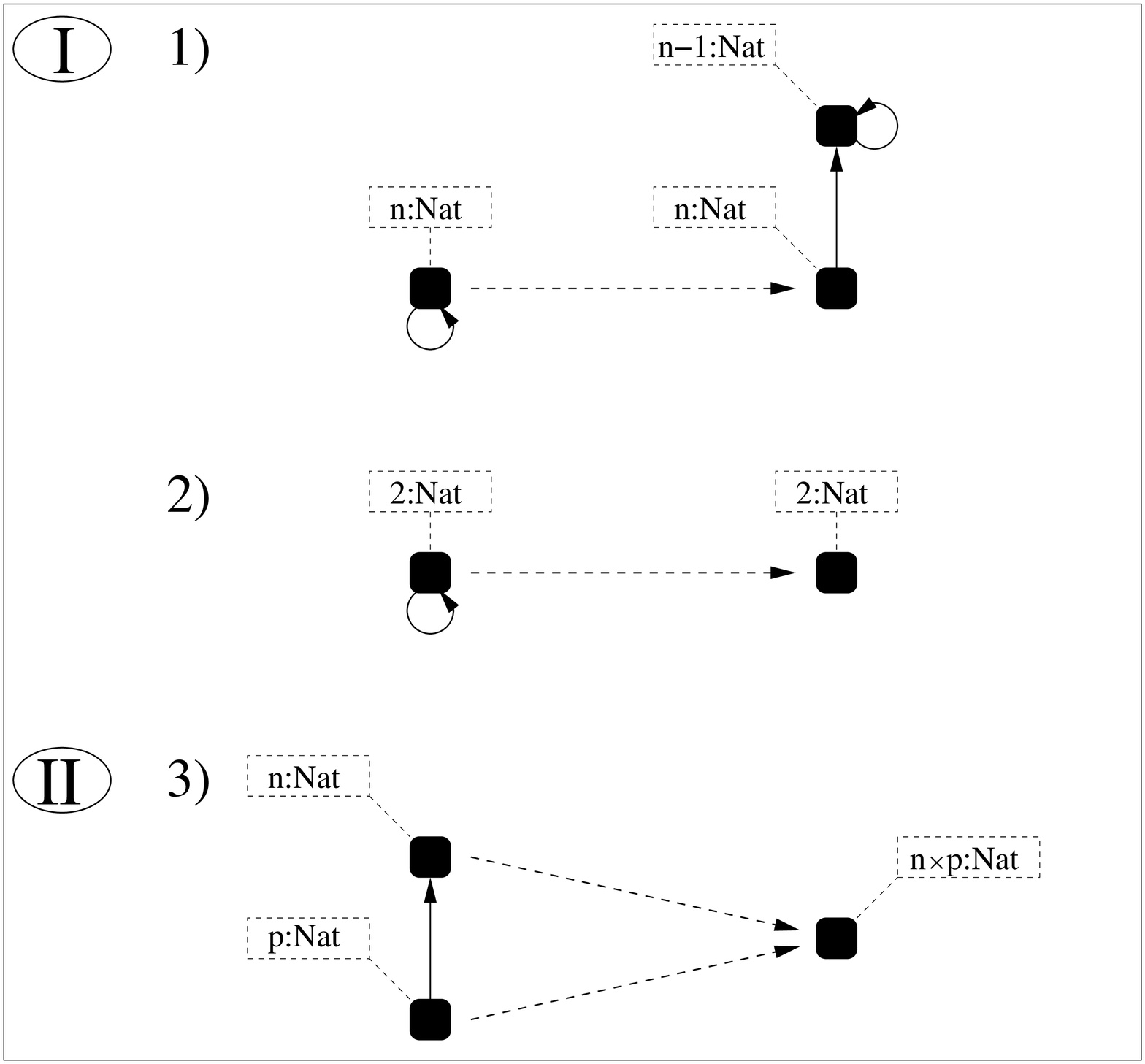}
    \label{fig:factoriellesigmaalgebras}
  }\qquad\qquad
  \subfigure[$n!$ computations based on $\lambda$-terms. Here we write ``$x!$'' instead of the $\lambda$-term computing $x!$ to lighten the figure.]{
    \includegraphics[width=0.41\textwidth]{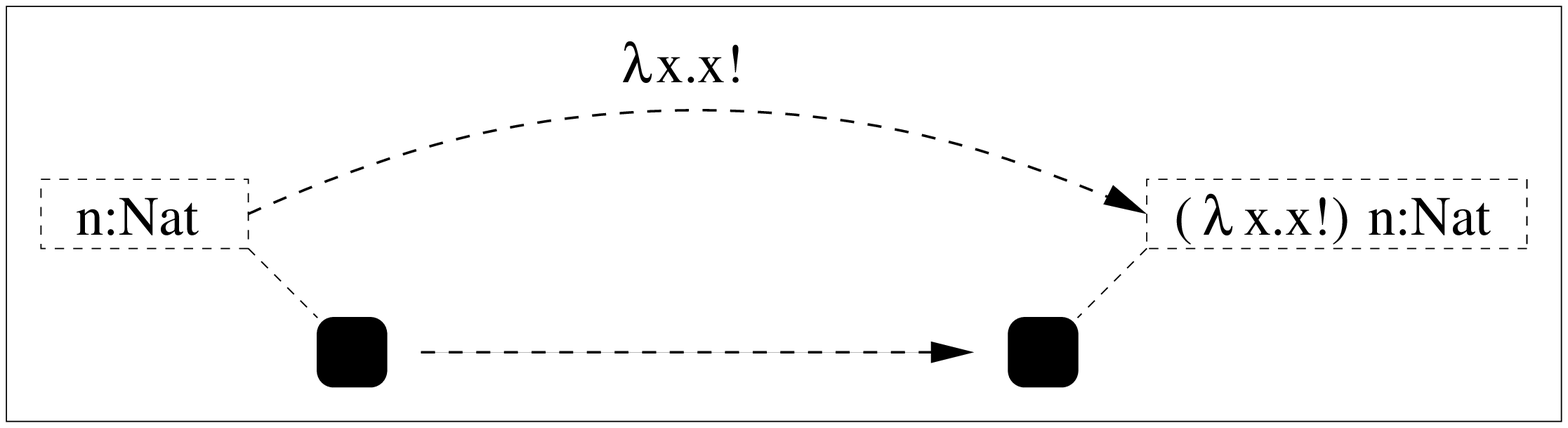}
    \label{fig:factoriellelambdaterms}
  }
  \caption{graph grammars to compute $n!$ in two different attributed graph rewriting systems}
  \label{fig:factorielle}
\end{figure}

This example shows the advantage of our approach at computational level. Figure \ref{fig:factorielle} presents the example of computing the factorial of a number $n$ using two different graph rewriting systems. The first one is based on $\Sigma$-algebras and is a copy of an example given on the AGG website \cite{AGG_homepage}, and the other is based on our approach using $\lambda$-terms.

If we use classical approach based on $\Sigma$-algebras, we need three rules in two layers (layers define a priority of application on the rules~\cite{fundamentals_alg_graph_transf} and are depicted by Roman ciffers on the figures) (cf figure \ref{fig:factoriellesigmaalgebras}):
\begin{enumerate}
\item the first rule is used $n-2$ times and creates a chain with all values between $n$ and $2$ (the looping edge is used to specify which attribute must be decremented);
\item the second rule is used to terminate the excecution of the first rule (it consists in removing the looping edge);
\item the third rule is used $n-2$ times to multiply all numbers between $n$ and $2$.
\end{enumerate}
Thus to compute $n!$ we must apply $2n-3$ rules.
This number of rule computations is due to the fact that in graph rewriting systems each rule can modelize the application of certain operations of a $\Sigma$-algebra, but putting to work recursion in $\Sigma$-algebras is difficult. Thus, the computation of a factorial is not straightforward.

If we use our framework based on $\lambda$-terms, only one rule is necessary to compute the factorial of $n$ because recursion operators exist, thus it is easy to write a $\lambda$-term that compute $n!$ (Cf figure \ref{fig:factoriellelambdaterms}). Of course computation of recursive functions requires many steps but it is included in a standard computational framework based on inductive types and optimized for such computations.

This reduction of the number of rules needed to compute on attributes has two advantages:
\begin{itemize}
\item the simplicity of the graph grammar;
\item the application of a rule is expensive in terms of algorithmic complexity. To apply a rule we need to solve the problem of subgraph isomorphism (wich is NP-complete) to find a matching. If we simplify the left side of rules and if we reduce the number of applications we significantly reduce the time of computing of $n!$.
\end{itemize}

Resting on this example, it appears that the attribute computation is more expressive and certainly more efficient than the one based on $\Sigma$-algebras, particularly when addressing problems requiring recursion.\\

\subsection{Managing infinity with functional attributes.}

\begin{figure}[h!]
  \centering
  \subfigure[Definition of the inductive type $\omega$-trees]{
    \begin{minipage}{20cm}
    \begin{tabbing}
      $T_\omega = Ind \alpha\{$\=$Leaf : \alpha,$\\
      \>$Succ_\omega: \alpha \rightarrow \alpha,$\\
      \>$Lim : (Nat \rightarrow \alpha) \rightarrow \alpha\}$
    \end{tabbing}
    \end{minipage}
    \label{fig:wTree1}
  }\qquad\qquad
  \subfigure[Transformation rule]{
    \includegraphics[width=0.4\textwidth]{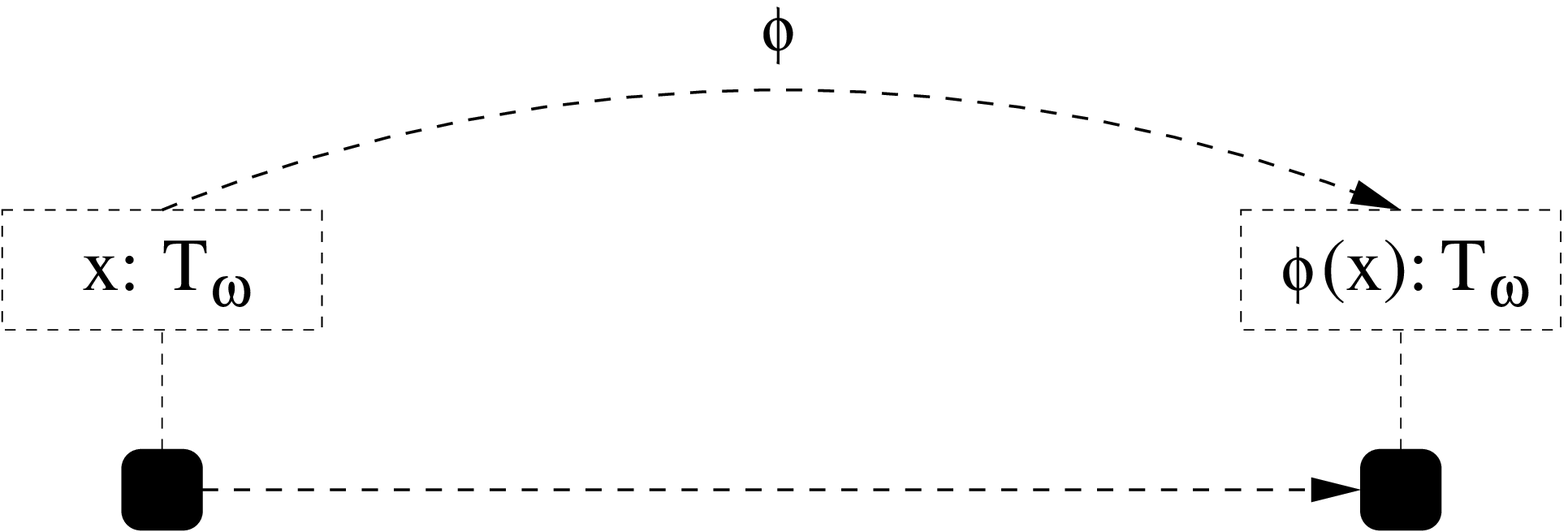}
    \label{fig:wTree2}
  }\qquad\qquad
  \subfigure[Example of $\omega$-tree defined by the term $Lim(Rec^{Nat\rightarrow T_{\omega}} (Leaf) (\lambda x^{Nat}\lambda y^{T_\omega}.Succ_{\omega}(y)))$. The length of the n-th branch is n.]{
    \includegraphics[width=0.3\textwidth]{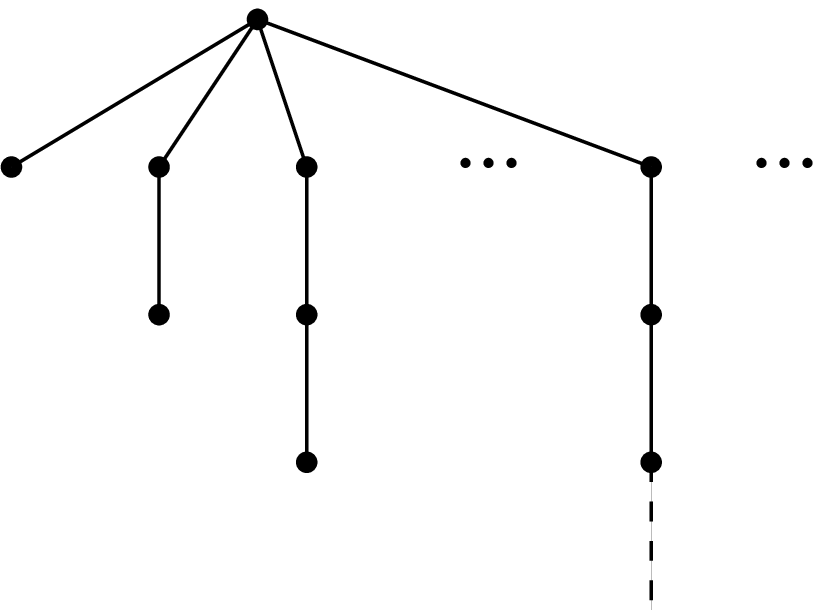}
    \label{fig:wTree3}
  }\qquad\qquad
  \subfigure[Result of the application of the rule \ref{fig:wTree2} on a graph with the $\omega-Tree$ of figure \ref{fig:wTree3} as attribute: only the branches with pair numbers are selected.]{
    \includegraphics[width=0.3\textwidth]{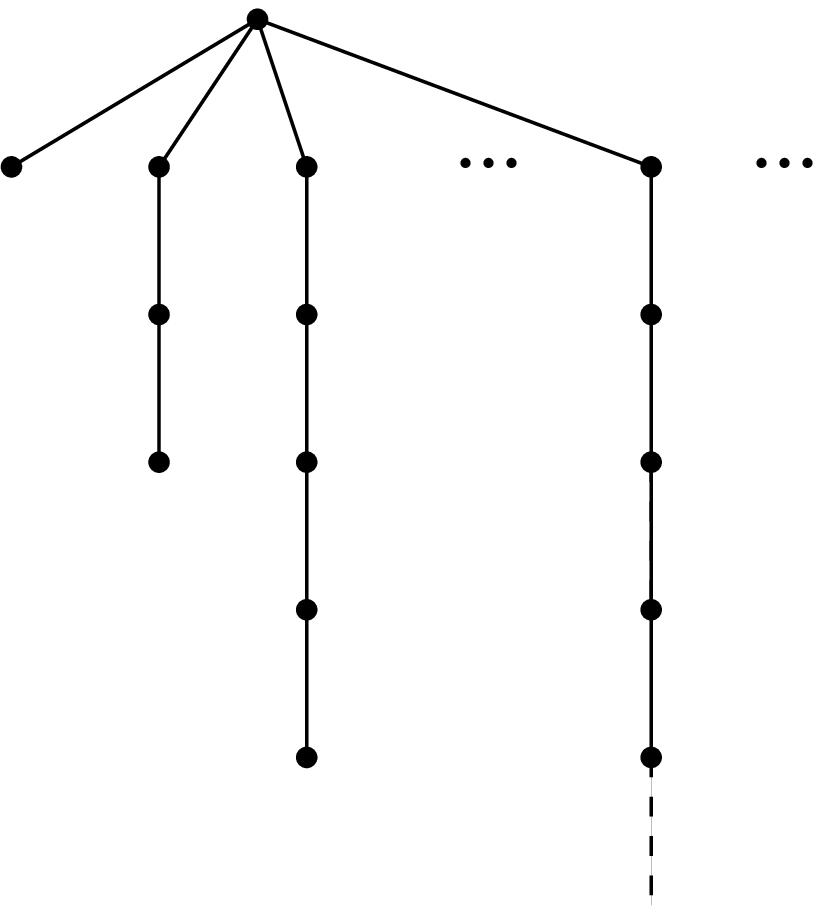}
    \label{fig:wTree4}
  }

  \caption{Computation on attributes representing infinite trees.}
  \label{fig:transfertAttributStructureAB}
\end{figure}

Another advantage of using $\lambda$-terms as attributes is the possibility to have complex data structures that can represent infinity. If we want to modelize infinite attributes we can do that by defining a function with an infinite data type as domain. For example the type $T_\omega$ of $\omega$-trees (cf figure \ref{fig:wTree1}) represents trees where nodes can have an infinity of subtrees. Using the recursion operators on inductive types we can define transformation on these infinite tree structures.

We may recall the form of recursive equations from $Nat$ to any type $A$ and from $T_\omega$ to $B$:
\begin{itemize}
  \item $\psi(0) = a; \psi(succ(x)) = g(x,\psi(x))$ (for Nat)
  \item and $\phi(Leaf) = t; \phi(Succ_\omega(x)=g(x, \phi(x)); \phi(Lim(f)) = h(f, \phi  \circ f)$ ($\circ$ denotes the composition).
\end{itemize}
In the syntax of $\lambda$-calculus the solution of these equation is denoted by
\begin{itemize}
\item $Rec^{Nat\rightarrow A}(a)(g)$
\item and $Rec^{T_\omega \rightarrow B}(t)(g)(h)$.
\end{itemize}
Now the transformation of the trees we are considering may be written as follows.
\begin{itemize}
\item Let $d$ be defined by $d(0) = 0, d(Succ(x))= Succ(Succ(d(x)))$, $d= Rec^{Nat\rightarrow Nat}(0)(\lambda x.\lambda y.Succ(Succ(y)))$, i.e. $d(x)=2x$ in arithmetical notation.
\item Let $\phi$ be defined by  $\phi(Leaf)=Leaf, \phi(Succ_\omega(x))= Succ(\phi(x)), \phi(Lim(f)) = Lim (f \circ d)$, using $Rec$, $\phi(f) = Rec^{T_\omega \rightarrow T_\omega} (Leaf) (\lambda x^{T_\omega}. Succ_\omega)(\lambda u.\lambda v.(u \circ d))$. This transformation selects (once) the branches with pair numbers at the first (infinite) branching.
\end{itemize}
Slightly more sophisticated transformation selects the branches with pair numbers at every infinite branching. There is only one modification.
We define $\phi'$ by  $\phi'(Leaf)=Leaf$, $\phi'(Succ_\omega(x))= Succ(\phi'(x))$, $\phi'(Lim(f)) = Lim ((\phi' \circ f) \circ d)$. Using $Rec$, $\phi'(f) = Rec^{T_\omega \rightarrow T_\omega} (Leaf) (\lambda x^{T_\omega}. Succ_\omega)(\lambda u.\lambda v.(v \circ d))$.

The figure \ref{fig:wTree2} presents a rule that selects the branches with pair numbers at the first infinite branching (using $\phi$ defined above). The figure \ref{fig:wTree3} presents an example of $\omega-tree$ and the figure \ref{fig:wTree4} presents the result of the application of the rule on it.

This is a very simple example but we can imagine a more complex example where leafs are elements of a complex type.

\section{Conclusion.}

The aim of this paper was to present a new attributed graph rewriting system based on the the SPo approach and whose main originality concerns the use of a typed $\lambda$-calculus to express attribute computations. On the structural parts our approach has the same characteristics than the classical SPo approach, but on the computation on attributes we have shown by examples that we can simplify the grammars, extend the expressivity of rules and certainly gain in efficiency of the computation.

Thanks to the expressive power of inductive types, it is now possible to dispatch some rewrite mechanism from structure to attribute computation and back (attributes can represent certain types of graphs e.g. trees).

Theoretically speaking, the SPo approach necessitates the definition and the construction of a weak pushout when dealing with attributes. A solution is presented in this paper. The next step of this work will concern the study of usual properies of any rewriting system such as confluence, termination, critical pairs analysis, etc. Note that for attribute computations these properties are well known properties of $\lambda$-calculus. In addition, we are now investigating another way to describe transformation of attributes, based on a calculus using deduction rules.

The possible domains of
applications include all usual applications of graph transformations,
e.g., verification and model transformations in programming,
but more ``tight'' relationship between computational and structural
parts will permit also the pursuit of much more specific goals.

\newpage
\nocite{*}
\bibliographystyle{eptcs}
\bibliography{biblio}
\end{document}